# Performance Evaluation of the Symmetrical Quasi-Classical Dynamics Method based on Meyer-Miller Mapping Hamiltonian in the Treatment of Site-Exciton Models

*Yu Xie[a], Jie Zheng[b] and Zhenggang Lan[*a,c]*

[a] Qingdao Institute of Bioenergy and Bioprocess Technology, Chinese Academy of Sciences, Qingdao, 266101 Shandong, China

[b] Industrial Research Institute of Nonwovens & Technical Textiles, College of Textiles & Clothing, Qingdao University, Qingdao 266071, China

[c] The Environmental Research Institute; MOE Key Laboratory of Theoretical Chemistry of Environment, South China Normal University, Guangzhou 510006, China

## Abstract

The symmetrical quasi-classical dynamics method based on the Meyer-Miller mapping Hamiltonian (MM-SQC) shows the great potential in the treatment of the nonadiabatic dynamics of complex systems. We performed the comprehensive benchmark calculations to evaluate the performance of the MM-SQC method in various site-exciton models with respect to the accurate results of quantum dynamics method multilayer multiconfigurational time-dependent Hartree (ML-MCTDH). The parameters of the site-exciton models are chosen to represent a few of prototypes used in the description of photoinduced excitonic dynamics processes in photoharvesting systems and organic solar cells, which include the rather board situations with the fast or slow bath and different system-bath couplings. When the characteristic frequency of the bath is low, the MM-SQC method performs extremely well, and it gives almost the identical results to those of ML-MCTDH. When the fast bath is considered, the deviations exist between the MM-SQC and ML-MCTDH results if the high-frequency bath modes are improperly treated by the classical manner. When the so-called adiabatic renormalization was employed to construct the reduced Hamiltonian by freezing high-frequency modes, the MM-SQC dynamics can give the results comparable to the ML-MCTDH ones. Thus, the MM-SQC method itself provide reasonable results in all test site-exciton models, while the proper treatments of the bath modes must be employed. The possible dependence of the MM-SQC dynamics on the different initial sampling methods for the nuclear degrees of freedom is also discussed.

# 1. INTRODUCTION

The theoretical understanding of nonadiabatic processes is a very challenging topic because the Born−Oppenheimer approximation breaks down due to the involvement of strongly coupled nuclear/electronic motions.[1-3] In last decades considerable efforts were made to develop various theoretical methods, from full quantum dynamics to different versions of semi-classical or mixed quantum classical dynamics, to solve the nonadiabatic dynamics of complex systems [1-41]. The full quantum dynamics methods[7,8] can give the very accurate description of the nonadiabatic dynamics, while the computational cost is extremely high even unaffordable in complex systems with a huge number of degrees of freedom. The Gaussian-wavepacket based methods,[12] such as multiple spawning,[11,42,43] variational multiconfigurational Gaussians (vMCG)[44-47] and multiconfigurational Ehrenfest,[48,49] may significantly improve the computational efficiency with keeping reasonable computational accuracy, while a significant amount of computational cost is still required in large systems and numerical implementation also required non-trivial tricks, for instance the proper generation of the new Gaussian wave-packets [43] or the reasonable initial samplings[50]. Several efficient mixed quantum classical approaches, such as Ehrenfest dynamics[20] and its extensions,[51] or trajectory surface-hopping dynamics[5,6,13,24,37-40], etc., are very practical in the treatment of the nonadiabatic dynamics of large systems. Thus, great efforts were made to examine the performance of these very efficient methods in various nonadiabatic processes [1,2,6,13,23,24,38-40,52-57]. At the same time, the on-the-fly versions of these methods were developed to simulate the nonadiabatic dynamics of realistic polyatomic systems including all nuclear degrees of freedom[6, 32, 36, 37, 58-63]. However, we should not neglect the deficiencies of these highly approximated methods, such as the improper treatment of electronic coherence and the appearance of frustrated hops in fewest-switches trajectory surface-hopping method[13, 24, 34, 38, 64].

Meyer and Miller [65] proposed an alternative approach for the theoretical treatment of nonadiabatic dynamics, which is based on the mapping from discrete quantum states to continuous classical variables. The approach was revisited by Stock and Thoss in the view of Schwinger's transformation [66]. It is also possible to construct the mapping Hamiltonian by other theoretical approaches[67-72]. In principle, the MM mapping approach defines a transformation to construct a mapping Hamiltonian with continuous variables, which may give the equivalent dynamical results with respect to those governed by the original Hamiltonian. Thus, this idea can be employed in different dynamics approaches. For instance, it is possible to combine it with semiclassical initial value representation and its extension [38, 73-84] quantum classical Louiville equation (QCLE)[85-90], path integral [91-97], surface hopping [98-101], centroid molecular dynamics (CMD)[102] and ring-polymer molecular dynamics (RPMD) [103-107].

The quasi-classical dynamics based on the MM mapping Hamiltonian (MM-QC) received considerable interests due to its simplicity [52, 65, 108-119]. Berne and coworkers[52] indicated that the MM-QC approach outperforms the mean-field dynamics in the description of nonradiative electronic relaxation processes. Stock and Müller [117, 118] introduced the zero-point-energy (ZPE) correction into the MM-QC approach to solve the ZPE leakage problem. Golosov and Reichman [119] proposed an systematic approach based on the Taylor expansion of the time-dependent observables around t=0. This idea not only improves the short time performance for the MM-QC dynamics, but also predicts the better results for the long-time dynamics. This approach also gives an alternative way to perform the ZPE correction.

In the quasiclassical dynamics procedure, the 'bin' technique has been widely used for the assignment of the final "quantum" states [67, 120-123]. This 'bin' technique was employed to assign the final electronic quantum number in the MM-QC dynamics [111, 124]. Cotton and Miller [125] proposed that it is more rigorous to perform the window tricks in both the initial sampling and the final assignment of the quantum states in the classical simulations. This gives the so-called the symmetrical quasi-classical dynamics method based on the MM mapping Hamiltonian (MM-SQC)[126, 127] in which the window width is also relevant to the ZPE correction of the mapping electronic degrees of freedom. In fact, this idea of the double "binning" average procedure appeared in Miller's earlier work. [128]

Recently, the MM-SQC method was received much attention and gained significant development. [54-56, 98-101, 126, 127, 129-139] Cotton and Miller [135] indicated that the MM-SQC method can obtain the correct results in Tully's scattering models [140] with anharmonicity. They also proved that the MM-SQC method with the rectangle windowing technique may recover the detailed balances in the model composed of quantum states and a classical bath, [129] while the Ehrenfest dynamics cannot give the correct thermal equilibrium.[129, 140] Cotton and Miller [132] proposed a new symmetric windowing technique, triangle windowing technique, for the weak-interstate-coupling cases where original rectangle windowing technique fails. The triangle windowing technique not only improve the performance of the MM-SQC dynamics in the weak-interstate coupling cases but also outperforms the rectangle windowing technique in general cases. Subotnik and coworkers employed different mixed quantum classical approaches, including the MM-SQC dynamics，to treat several typical models, such as harmonic and anharmonic models, [56] bilinear harmonic models [55] and two-level systems coupled with electromagnetic fields [54]. They indicated that the MM-SQC method can provide a good description of detailed balance for models with a harmonic classical bath at a given temperature, while it cannot recover the correct equilibrium populations for models involving very harsh anharmonicity [56] that brings the numerical instability associating with "negative forces" [38, 91, 141-143]. They observed that the performance of the MM-SQC method in moldels with very harsh anharmonicity can be improved by using the very-narrow-width window.[56] Cao, Geva and coworkers [131] modified the MM-SQC method to treat the anharmonic systems by including the vibrational levels of the anharmonic nuclear modes directly in the mapping procedure. By using this approach, the modified MM-SQC method can accurately describe the population dynamics of systems involving very harsh anharmonic potential energy surfaces (PESs). Tao developed a few of the extended versions of the MM-SQC method by combining the MM-SQC method with the trajectory surface hopping. [98-101] At the same time, the MM-SQC approach has been applied to treat several typical situations [126, 127, 133-139], such as electron transfer[126, 133], energy transfer[134, 139], singlet fission[136-138], etc.

Overall, the MM-SQC method is a promising method to treat the nonadiabatic dynamics, thus it is highly interesting to put additional efforts to apply this idea to investigate the nonadiabatic dynamics of more complex systems. Before the formal implementation for the large-scale simulation, it is also necessary to perform the extensive benchmark calculations to give more comprehensive evaluations on the performance of the MM-SQC method. For this purpose, we try to examine the accuracy of the MM-SQC method in the dynamics of the site-exciton models used to describe the electron transfer and energy transfer processes in complex systems widely[3, 4]. Particularly, we wish to focus on the performance of MM-SQC dynamics in the site-exciton models with different parameters, such as energy gap, electronic couplings, the bath characteristic frequency and electron-phonon coupling. As a benchmark, the full quantum dynamics method, ML-MCTDH[8, 144, 145], was used to obtain the accurate dynamics results of these model systems for comparison. This work serves as an important complement to the available studies of the MM-SQC method[54-56, 98-101, 126, 127, 129-138, 146, 147], before the massive employment of this approach in the treatment of more general nonadiabatic dynamics. This work definitely improves our understanding on the performance of the MM-SQC method. Because the site-exciton models with quite different parameters were considered, the current work will give us the direct information on the general performance of the MM-SQC method. This is certainly useful for the further studies on similar problems.

## 2. THEORY AND METHODS

**2.1 Hamiltonian**

We employed the site-exciton models in this paper, which were widely used to describe the excitonic dynamics in previous works[4, 134, 139, 148]. Here we mainly consider the site-exciton models that describe the typical situations of two electronic states (system) coupled with many vibrational modes (bath). A system−bath Hamiltonian including three parts was used to describe the dynamics, namely: the system Hamiltonian $H_s$ (electronic part) the bath part $H_b$ (vibrational part) and the

system−bath couplings $H_{sb}$ (electronic-phonon interaction); namely,

$$H = H_s + H_b + H_{sb}, \tag{1}$$

Each part can be further expanded into the following form,

$$\begin{aligned}
H_s &= \sum_k |\phi_k\rangle E_k \langle\phi_k| + \sum_{l\neq k} |\phi_k\rangle V_{kl} \langle\phi_l| \\
H_b &= \sum_k \sum_j^{N_b} \frac{1}{2}\omega_{kj}\left[P_{kj}^2 + Q_{kj}^2\right] \\
H_{sb} &= \sum_k |\phi_k\rangle\langle\phi_k| \sum_j^{N_b}\left(-\kappa_{kj}Q_{kj}\right)
\end{aligned} \tag{2}$$

where $\phi_k$ and $\phi_l$ represent the electronic states and $E_k$ is the energy of the $k$-th state. $V_{kl}$ is the interstate coupling. Each state couples with an individual bath. $P_{kj}$ and $Q_{kj}$ are the momentum and position of the $j$-th mode in the $k$-th bath coupled with the $k$-th state. And $\omega_{kj}$ and $\kappa_{kj}$ are the frequency and the system-bath coupling constant of the corresponding mode in the diagonal elements of the system Hamiltonian, respectively. $N_b$ is the bath-mode number. The system−bath couplings for the off-diagonal elements in the system Hamiltonian were neglected. Four different models with various interstate energy gaps and couplings were adopted using the above Hamiltonian.

Here, the continuous Debye-type spectral density[3] was used for bath modes,

$$J(\omega) = \frac{2\lambda\omega\omega_c}{\omega^2 + \omega_c^2}, \tag{3}$$

where $\omega_c$ is the characteristic frequency of the bath. $\lambda$ is the reorganization energy representing the coupling strength between system and bath degrees of freedom. Each electronic state was coupled to its corresponding bath. Discrete bath modes were also used to define the spectral density[3]

$$J_k(\omega) = \frac{\pi}{2}\sum_{i=1}^{N}\kappa_{ki}^2 \delta(\omega-\omega_{ki}). \tag{4}$$

In principle, the continuous spectral density can be discretized to an arbitrary number of bath modes. The coupling coefficient $\kappa_{ki}$ of each mode is obtained,

$$\kappa_{ki} = \sqrt{\frac{2}{\pi}J_k(\omega_{ki})\Delta\omega}, \tag{5}$$

when a sampling interval $\Delta\omega$ was given. The coefficient $\kappa_{ki}$ stands for the coupling strength of the mode. Normally the electronic-phonon couplings are well characterized by the reorganization energy, or the values of the coupling strength divided by the frequency of the corresponding mode. In the discretization of the bath modes, different bath mode number gives different coupling strength $\kappa_{ki}$ for the modes in the same frequency domains. Thus, it is not straightforward to employ the coupling strength to represent the system-bath couplings in the current situation. Instead, we employed a rather simple and direct approach here. Let us assume that the bath organization energy is represented by the single mode with the characteristic frequency, then we can define the effective parameter $\kappa/\omega_c$,

$$\kappa/\omega_c = \sqrt{\frac{2\lambda}{\omega_c}}, \tag{6}$$

to characterize electron-phonon coupling strength of the bath.

## 2.2 Symmetry-Quasi-Classical Dynamics Method Based on Mapping Hamiltonian

The $\hat{H}$ is used as Hamiltonian of a $N$-state system

$$\hat{H} = \sum_{k,l}\hat{h}_{kl}|\phi_k\rangle\langle\phi_l|, \tag{7}$$

where $\phi_k$ and $\phi_l$ denote the quantum states of the system. The diagonal element $\hat{h}_{kk}$ represents the energy operator (kinetic and potential energies). The off-diagonal element $\hat{h}_{kl}(k \neq l)$ denotes the interstate coupling operator. This Hamiltonian is represented in the discrete quantum-state basis ($\phi_k$). The main idea of the mapping approach is to map the discrete representation to a continuous one.

In the MM model[66, 127], a $N$-state system is mapped to a $N$ coupled harmonic oscillators. The annihilation operator $\hat{a}_k$ and the creation operator $\hat{a}_l^+$ were used to construct the mapping relation, namely

$$|\phi_k\rangle\langle\phi_l| \mapsto \hat{a}_k^+ \hat{a}_l,$$
$$|\phi_k\rangle \mapsto |0_1 \cdots 1_k \cdots 0_N\rangle. \quad (8)$$

The relationship between the annihilation operator and the creation operator is given as $\left[\hat{a}_k, \hat{a}_l^+\right] = \delta_{kl}$. When the Cartesian coordinate operator $\hat{x}_k = \left(\hat{a}_k^+ + \hat{a}_k\right)/\sqrt{2}$ and the momentum operator $\hat{p}_k = i\left(\hat{a}_k^+ - \hat{a}_k\right)/\sqrt{2}$ of the harmonic oscillator were introduced, the MM Hamiltonian $\hat{H}$ is written as

$$\hat{H} = \sum_k \left[\frac{1}{2}\left(\hat{x}_k^2 + \hat{p}_k^2\right) - \frac{1}{2}\right]\hat{h}_{kk} + \frac{1}{2}\sum_{k \neq l}\left(\hat{x}_k\hat{x}_l + \hat{p}_k\hat{p}_l\right)\hat{h}_{kl}, \quad (9)$$

A purely classical mapping Hamiltonian can be obtained by substituting the quantum operators with their corresponding physical variables in the MM Hamiltonian, namely

$$H_{MM}(x,p,Q,P) = \sum_k \left[\frac{1}{2}\left(x_k^2 + p_k^2\right) - \gamma\right]H_{kk}(Q,P) + \frac{1}{2}\sum_{k \neq l}\left(x_k x_l + p_k p_l\right)H_{kl}(Q,P), \quad (10)$$

where $x_k$ and $p_k$ are the classical coordinates and momenta for the quantum-state part. $Q$ and $P$ are the coordinates and momentums of the environmental part. $H_{kk}$ and $H_{kl}$ are the classical correspondences of the quantum Hamiltonian matrix elements. $\gamma$ is a parameter accounting for the effective zero-point energy. It is possible to perform the proper initial sampling and run the classical dynamics over many trajectories starting from this classical mapping Hamiltonian, which provides a quasi-classical trajectory-based approach to treat the nonadiabatic dynamics.

In the quasi-classical dynamics, the distribution of final quantum states (occupation) can be obtained by calculating the final values of the action variables $n_k = \frac{1}{2}\left[\left(x^{(k)}\right)^2 + \left(p^{(k)}\right)^2\right] - \gamma$ accumulated in square histograms ("bin") centered at quantum occupation values $N$ = 0 or 1. Cotton and Miller proposed that the "bin" idea should be employed for both initial-state sampling and final-state assignment[127], giving the so-called symmetrical quasiclassical (SQC) dynamics. In the SQC method, the "window functions" $w_k$ is defined as

$$w_k(n_k, N) = \frac{1}{2\gamma} h\left(\gamma - |n_k - N|\right), \tag{11}$$

where $h(z)$ is the Heaviside function, $h(z)$ = 0 when $z$ < 0 and $h(z)$ = 1 when $z \geq 0$. The window width is recommended to be $2\gamma$ in the MM-SQC method.[135] The parameter $\gamma$ is generally chosen as 0.366 for rectangle window function.[135] For an arbitrary $F$ number of electronic states, a joint window function with $k$-th electronic state is occupied may be written as

$$W_k(\mathbf{n} = n_1, ... n_k, ... n_F) = \frac{1}{(2\gamma)^F} h\left(\gamma - |n_k - 1|\right) \prod_{k' \neq k}^{F} h\left(\gamma - |n_{k'}|\right), \tag{12}$$

In the MM-SQC dynamics simulations, the initial-state sampling can be performed using the action-angle sampling method, namely

$$x_k = \sqrt{2(n_k + \gamma)} \cos\theta$$
$$p_k = \sqrt{2(n_k + \gamma)} \sin\theta \quad , \qquad (13)$$

where $n_k \in [N - \gamma, N + \gamma]$ ($N = 0$ or $1$) and the angle $\theta \in [-\pi, \pi)$. For the final states, the same binning way is also performed for the assignment of the quantum state.

Recently Cotton and Miller proposed the triangle window function. For a 2-state model, such functions are defined as

$$W_1(n_1, n_2) = 2 \cdot h(n_1 + \gamma - 1) \cdot h(n_2 + \gamma) \cdot h(2 - 2\gamma - n_1 - n_2)$$
$$W_2(n_1, n_2) = 2 \cdot h(n_1 + \gamma) \cdot h(n_2 + \gamma - 1) \cdot h(2 - 2\gamma - n_1 - n_2) \quad . \qquad (14)$$

Equation (14) gives two triangle windows. The parameter $\gamma$ is generally chosen as 1/3 for triangle window function. For further details, please refer to the work by Cotton and Miller [132].

For both types of window functions, the time-dependent populations $P_{map,k}$ of electronic states are evaluated by averaging over all trajectories

$$P_{map,k}(t) = \frac{\langle W_k(\mathbf{n} = n_1, ...n_k, ...n_F) \rangle}{\sum_m^F \langle W_m(\mathbf{n} = n_1, ...n_m, ...n_F) \rangle} \qquad (15)$$

## 2.3 ML-MCTDH

In the traditional quantum wavepacket dynamics, a group of time-independent bases for each primary coordinate is used to represent the wavefunction, only the time-dependent coefficients evolving with time. The MCTDH method, instead, employs the time-dependent bases to represent wavefunction. The equations of motion of the coefficients and the basis functions can be obtained by the variational principle [7].

The ML-MCTDH approach[8, 144, 145] is an extension of the standard MCTDH method. The wavefunction is represented by a multilayer tree structure, which is expanded by the multi-dimensional basis functions recursively until the time-independent bases of each primary coordinate are reached. In principle, the ML-MCTDH provides the numerically accurate simulation of the quantum dynamics of complex systems. The expansion of tree structure has a great influence on the computational convergence and efficiency. The principle for the construction of tree structure has been discussed in detail in the previous works [4, 148-151], which will not be restated here. In this work, ML-MCTDH calculations were conducted to get the accurate dynamical results for benchmark.

**2.4 Adiabatic Renormalization**

Suppose a bath contains both low-frequency and high-frequency bath modes. For the later ones, their frequencies may be much larger than the diabatic coupling. This gives the possibility to use the adiabatic (Born-Oppenheimer) type of approximation to treat these modes. This give the so-called "adiabatic renormalization" model.[152-158]. Please notice that in the current site-exciton model, the motion of some vibrational modes is much faster than the electronic motion. This situation is reversed to the normal BO approximation in the chemical dynamics.

The detailed derivation of the adiabatic renormalization model were given in previous work[152, 153] and such idea was also discussed widely[154-158], we thus only outline its theoretical framework briefly and focus on its application in our current work. Formally, the "adiabatic renormalization" model assumes that all high-frequency modes (with frequencies higher than a cutoff value $\omega_{cut}$) are frozen into the ground vibrational level of each diabatic electronic state, when the diabatic coupling is significantly smaller than $\omega_{cut}$. After we put the electronic wavefunction of the site excitons and the ground vibrational wavefunctions of these high-frequency modes together, it is possible to redefine the vibronic levels with the below new interstate coupling

$$V'_{12} = V_{12} \prod_i^{\omega_i > \omega_{cut}} \int_{-\infty}^{+\infty} \chi_0\left(Q_i - \Delta Q_{i,0}^{(1)}\right) \chi_0\left(Q_i - \Delta Q_{i,0}^{(2)}\right) dQ_i, \tag{16}$$

where $V_{12}$ is the original electronic coupling between the two electronic states and $V'_{12}$ is the rescaled inter-state coupling. $\chi_0\left(Q_i - \Delta Q_{i,0}^{(\alpha)}\right)$ ($\alpha$ = 1, 2) denotes the ground state of the harmonic oscillator of the $i$-th mode on the potential surface of State $\alpha$. $Q_i$ is the coordinate of the $i$-th mode. $\Delta Q_{i,0}^{(\alpha)}$ is the coordinate shift from the electronic ground-state minimum to the State-$\alpha$ minimum. This shift is expressed as $\Delta Q_{i,0}^{(\alpha)} = \kappa_i^{(\alpha)}/\omega_i$, where $\omega_i$ are the frequency of the $i$-th mode and $\kappa_i^{(\alpha)}$ is the corresponding vibronic coupling of the $i$-th mode to State $\alpha$. The product of Equation (16) runs over all the modes with frequencies higher than $\omega_{cut}$. In this way, a new model is obtained, in which we only need to deal with a reduced model that contains the low-frequency bath modes and shows the rescaled diabatic coupling (inter-state coupling, Eq. 16).

In the current work, each state is coupled to its own bath. For instance, when the

*i*-th bath mode is coupled to State 1, $\Delta Q_{i,0}^{(1)} = \Delta x_i$ and $\Delta Q_{i,0}^{(2)} = 0$; or vice versa. The $\Delta x_i$ value can be obtained using $\Delta x_i = \kappa_i / \omega_i$. Because two baths take the same spectral density, the rescaled inter-state coupling becomes

$$V'_{12} = V_{12} \prod_i^{\omega_i^{(1)} > \omega_{cut}} \int_{-\infty}^{+\infty} \chi_0\left(Q_i^{(1)} - \Delta x_i^{(1)}\right) \chi_0\left(Q_i^{(1)}\right) dQ_i^{(1)} \prod_i^{\omega_i^{(2)} > \omega_{cut}} \int_{-\infty}^{+\infty} \chi_0\left(Q_i^{(2)}\right) \chi_0\left(Q_i^{(2)} - \Delta x_i^{(2)}\right) dQ_i^{(2)} , (17)$$

where $Q_i^{(1)}$ is the coordinate of the *i*-th mode of the bath coupled to State 1 and $Q_i^{(2)}$ is the coordinate of the *i*-th mode of the bath coupled to State 2. In practices, the values of $\omega_{cut}$ and $V'_{12}$ are obtained using an iterating procedure, see the previous work [152] in details. We chose $\omega_{cut} = 5 V'_{12}$ in this work.

### 2.5 Computational Details

Since we built the models for charge and energy transfer problems, we name the initial state as the donor state in the whole context. In the current site-exciton model, the second state is the acceptor state. Here we also assume that both electronic states are excited states, and each state is coupled to an individual bath. According to the Condon approximation, the initial state is obtained by placing the lowest vibrational level of the ground electronic state into the donor state in both the ML-MCTDH and MM-SQC calculations. Please notice that the current initial condition gives the non-equilibrium nuclear density for the donor electronic state. The ML-MCTDH calculations were performed using the Heidelberg MCTDH package.[159]

In the MM-SQC calculations, the ZPE correction for the electronic degrees $\gamma = 0.366$ [135] for rectangle window functions and $\gamma = 1/3$ for triangle window functions were considered.

For the nuclear degrees of freedom, the Wigner sampling was used in most cases. In order to check the dependence of the MM-SQC dynamics on the different initial samplings of the nuclear degrees of freedom, the initial sampling for the bath modes was also conducted by using the action-angle sampling method with various ZPE corrections and window widths in a few of additional calculations.

## 3. RESULTS

Four groups of the site-exciton models with various energy gaps and inter-state couplings were considered in this work. The first two groups of the site-exciton models with small inter-state coupling (0.0124 eV) mimic the situations of the photoinduced energy transfer in photoharvesting systems [134, 160]. To check the performance of the MM-SQC dynamics in symmetric and asymmetric systems, the inter-state energy gap is chosen to be 0 and 0.0124 eV for the first and the second groups of the site-exciton models, respectively. The other two groups of the site-exciton models take large inter-state coupling 0.1 eV that is also a typical one in the photoinduced excitonic dynamics in organic solar cells [4, 151, 161-163]. Two inter-state energy gaps, 0.1 and 0.2 eV, are considered in the third and the fourth groups of the site-exciton models.

We took baths with the different characteristic frequencies and the various electron-phonon coupling strengths to check the performance of the MM-SQC method with respect to the ML-MCTDH results as benchmarks. The characteristic frequencies $\omega_c$ of the baths cover the range from 40 to 3000 cm$^{-1}$ and the parameter $\kappa/\omega_c$ characterizing electron-phonon coupling strength of the bath is chosen as 0.3, 0.5, 0.7 or 1.0.

The main purpose of this work to compare the results obtained by both the ML-MCTDH and MM-SQC calculations. As long as the same Hamiltonian is used in both two dynamical calculations, such comparison becomes meaningful. The current work employed 100 modes to represent the continuous bath. In this way, both two dynamical methods employed the identical Hamiltonian. This fully satisfies the purpose of the current work. In this sense, we do not worry whether the current discretization gives the enough number of the bath modes to represent the true bath or not, because this does not affect the main results of the current work.

**3.1 Symmetric Site-Exciton Models**

The first group of the site-exciton models included several symmetric ones with $\Delta E = 0$ and $V_{12} = 0.0124$ eV. The results of MM-SQC and ML-MCTDH are shown in Figure 1. When the characteristic frequency is low and the electron-phonon coupling is weak, we see the clear oscillation pattern in the time-dependent population dynamics. Such oscillation is governed by the electronic coherent dynamics and its amplitude decays slowly with time being. When the characteristic frequency of the bath increases or the electron-phonon coupling gets stronger, the oscillation amplitude of the time-dependent electronic population in the ML-MCTDH dynamics becomes weaker. As expect, in the cases with strong electron-phonon couplings and the high characteristic frequencies of the bath, the population oscillation vanishes and a monotonic population decay is observed.

In previous works[164], the ratio between the electronic coupling and the characteristic frequency of the bath $V_{12}/\omega_c$ was used to classify the electron transfer reactions into three regions, i.e., adiabatic (slow bath) region ($V_{12}/\omega_c \gg 1$), nonadiabatic (fast bath) region ($V_{12}/\omega_c \ll 1$) and intermediate region ($V_{12}/\omega_c \approx 1$).

In the adiabatic and intermediate bath regions ($\omega_c$ = 40, 100, 200 cm$^{-1}$ and $V_{12}/\omega_c$ <=2), the MM-SQC method give results very close to the ML-MCTDH ones, and even in some cases the identical results are obtained (Figure 1). Here in the MM-SQC model, all bath modes are treated purely by the classical dynamics.

In the nonadiabatic bath region, the MM-SQC results deviate from the ML-MCTDH ones, when all bath modes are treated by classical evolution. The population recurrences become weaker and vanishes more quickly in the MM-SQC calculations than those in the ML-MCTDH results in the weak electron-phonon coupling ($\kappa/\omega_c$ = 0.3 and 0.5) models. With intermediate electron-phonon coupling ($\kappa/\omega_c$ = 0.7), the MM-SQC dynamics display the monotonic decay patterns instead of the weak population oscillations in the ML-MCTDH dynamics. In the strong electron-phonon coupling ($\kappa/\omega_c$ = 1.0) models, the monotonic population decay in the MM-SQC dynamics is slower than that in the ML-MCTDH dynamics. The quicker damping of the population oscillation also appears in the previous study of nonradiative electronic relaxation by the MM-QC (without windowing) dynamics[52]. Moreover, the discrepancy between the ML-MCTDH and MM-SQC results becomes larger with the electron-phonon coupling increasing in the nonadiabatic bath region. The dependence of the accuracy of the MM-SQC dynamics on the Hamiltonian was also discussed by Subotnik and coworkers[55, 56]. In fact, the deviation of the MM-SQC results with respect to the ML-MCTDH ones is fully due to the improper treatment of the fast bath modes classically. When the proper treatment of the bath modes is employed, the MM-SQC method can give very reasonable results, see the below discussions in Section 3.5.

The previous work[132] indicated that the triangle window technique not only provides the better description of the weak electronic-coupling problems but also outperforms rectangle window method in normal cases. Thus, we calculated all testing models using both the rectangle window and the triangle window. In some cases, such as the models with $\omega_c$ = 200 cm$^{-1}$ and $\kappa/\omega_c$ = 0.5, 0.7, it is true that the triangle window method performs slightly better than the rectangle window method, see Figure 1 and Appendix I. In other cases, two different window functions give very similar results.

It is well known that the triangle window method works much better in the weak interstate coupling region, for instance for the two-site-exciton model with the electronic coupling of less than ~ 0.00124 eV [132]. However, when a larger electronic coupling is employed for the same site-exciton model, such as ~ 0.0124 eV, Cotton and Miller showed that the rectangle window also works very well. [134] Thus, we expect that the good performance of the rectangle window in the current work may be attributed to the fact that the interstate coupling in all above models (also ~ 0.0124 eV) may not be extremely small. However, we also noticed that the triangle window method gives better results in some models, although in other models both triangle and rectangle window methods work equally well. Thus, it is highly recommended to use the triangle window method in the MM-SQC method.

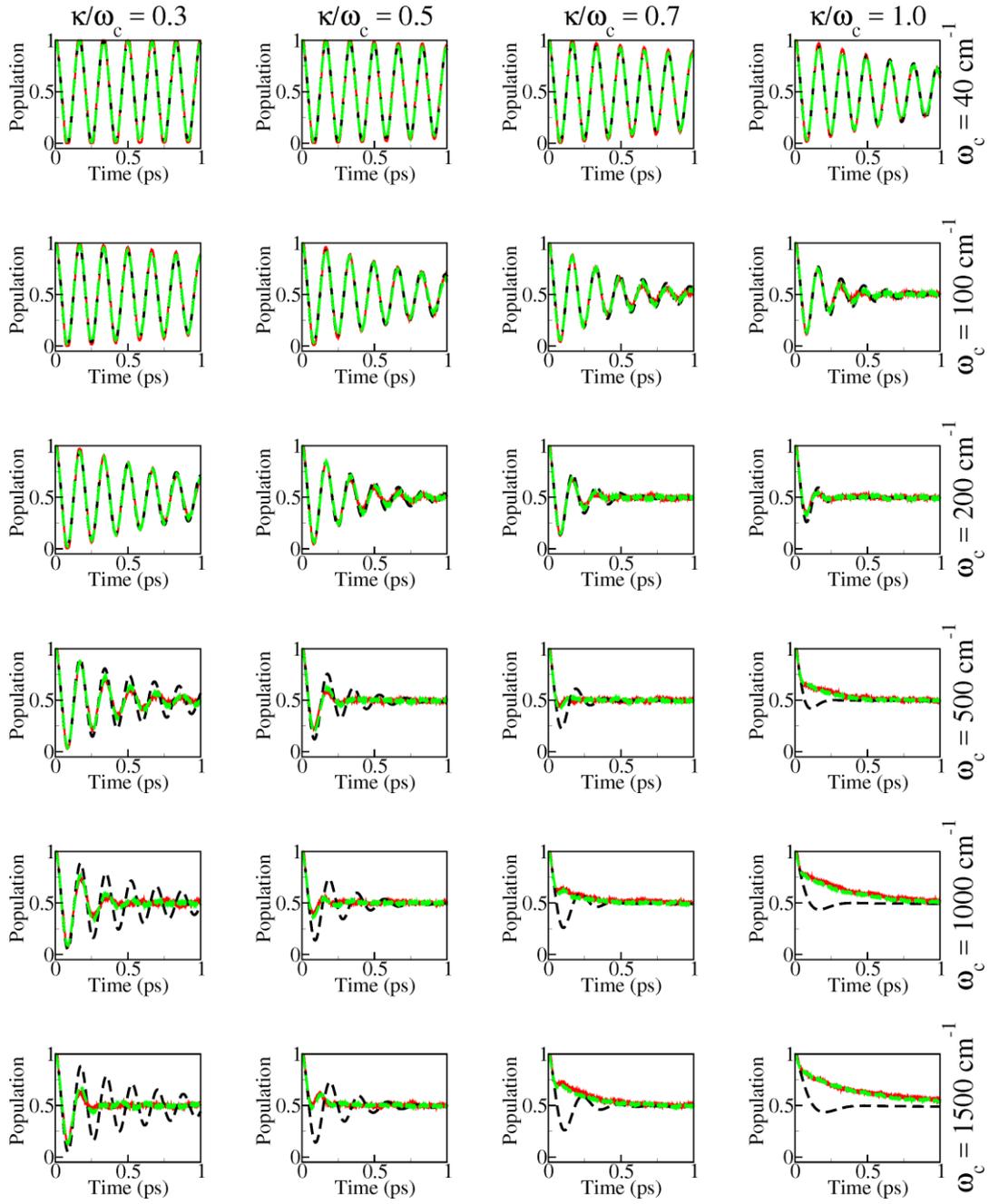

Figure 1. Time-dependent electronic population (or occupation) of the donor state in various two-state models with $\Delta E = 0$ and $V_{12} = 0.0124$ eV. The characteristic bath frequencies $\omega_c$ are 40, 100, 200, 500, 1000 and 1500 cm-1 and the electron-phonon coupling strength of the baths, $\kappa/\omega_c$, are 0.3, 0.5, 0.7 and 1.0. Black dashed lines denote the ML-MCTDH results. Red solid lines and green dashed lines denote the MM-SQC results using rectangle window and triangle window, respectively.

## 3.2 Asymmetric Site-Exciton Models with Weak Diabatic Couplings

A series of asymmetric models with the same parameters as the symmetric models above were considered except the energy gap $\Delta E$ = 0.0124 eV. The dynamical results are shown in Figure 2. In the ML-MCTDH results, the overall population evolution of the donor state in the asymmetric models is very similar to that in the symmetric models, except that different population plateaus are reached at the end. Because of the higher energy of the initial donor state, the final population of the donor state tends to be less than 0.5 at the end of the ML-MCTDH simulation.

Because all modes here do not use extremely small electronic couplings, the triangle window method also gives slightly better results in some models (see Figure 2 and Appendix I) and equally good results in the other models, compared to the rectangle window method. In the nonadiabatic bath region, the deviations exist between two different dynamics calculation results when the fast bath modes are treated classically.

The MM-SQC results are also generally consistent with the ML-MCTDH ones in the adiabatic and intermediate bath regions for these asymmetric models. In the models with rather high characteristic frequencies of the bath and strong electron-phonon couplings, the populations do not match at the end of simulation time.

Tully and coworkers pointed out that the standard Ehrenfest method cannot recover the detailed balance based on the model of a quantum oscillator coupled to a classical bath at a given temperature. [140] Recently, Miller and Cotton [129] proved that the MM-SQC method can provide a good description of detailed balance in the similar model. Subotnik and coworkers [56] indicated that the detailed balance is indeed achieved for the spin-boson model with a classical bath using the MM-SQC method, while it is not necessary to guarantee that the detailed balance is recovered in the classical treatment of the quantum bath motion in the MM-SQC dynamics. In addition, they also noticed several interesting features of the MM-SQC dynamics. For instance, the MM-SQC method with the window width of 0.366 performs well in the description of the equilibrium population for the spin-boson models with the weak diabatic coupling, while the very narrow window should be used in the treatment of the model in the present of strong anharmonicity.

In fact, the deviation of the MM-SQC dynamics in the fast bath cases is relevant to the improper treatment of the high-frequency bath modes by the classical dynamics. It is possible to remedy such problem by introducing the so-called "adiabatic renormalization" [152, 153] procedure and the results can be found in Section 3.5.

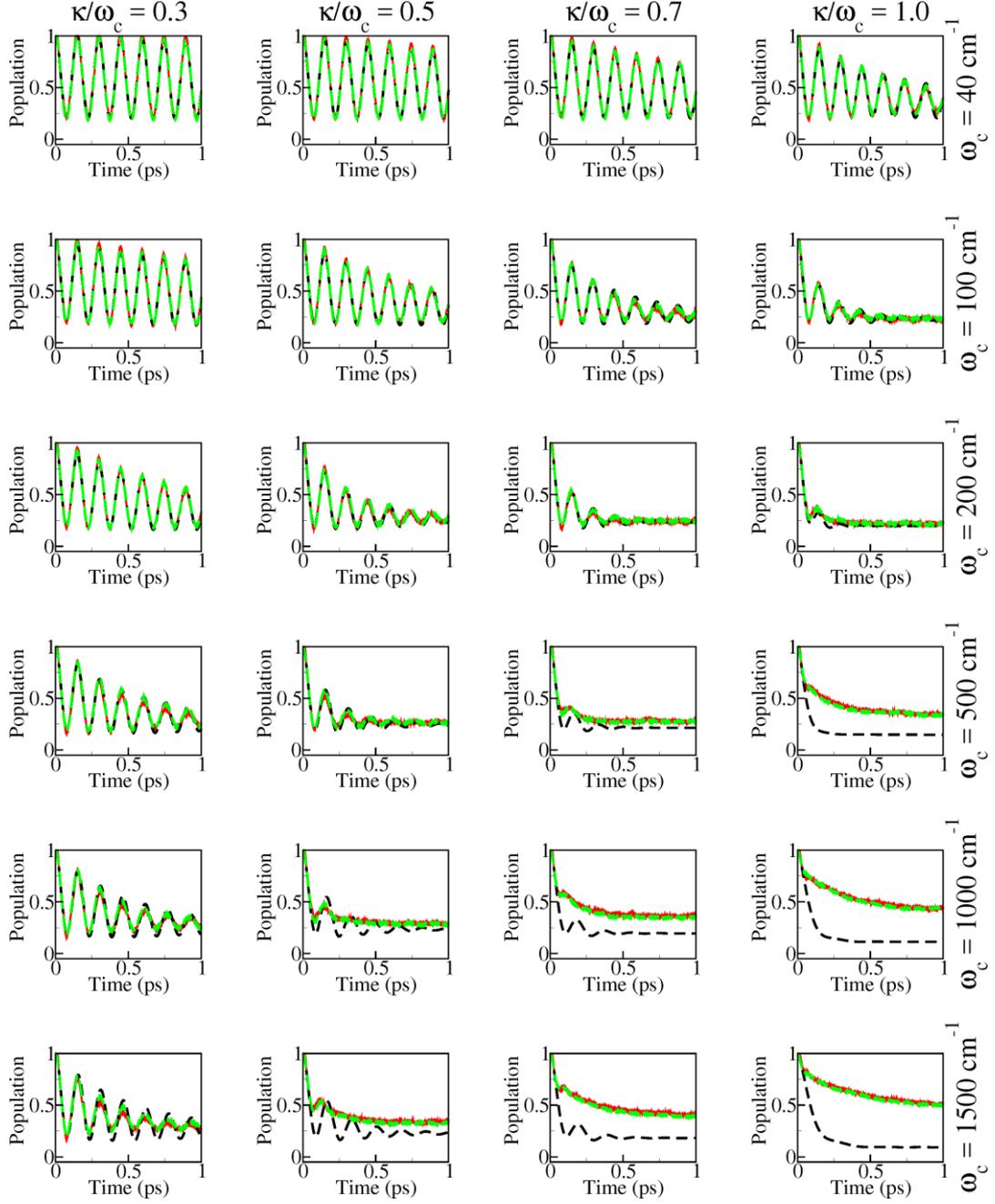

Figure 2. Time-dependent electronic population (or occupation) of the donor state in various site-exciton models with $\Delta E = V_{12} = 0.0124$ eV. The characteristic bath frequencies $\omega_c$ are 40, 100, 200, 500, 1000 and 1500 cm-1 and the electron-phonon coupling strength of the baths, $\kappa/\omega_c$, are 0.3, 0.5, 0.7 and 1.0. Black dashed lines denote the ML-MCTDH results. Red solid lines and green dashed lines denote the MM-SQC results using rectangle window and triangle window, respectively.

### 3.3 Asymmetric Site-Exciton Models with Strong Diabatic Couplings

Previous models with small inter-state coupling takes parameter values that were used in the discussions of photoinduced intermolecular energy transfer process in photoharvesting systems[134, 160]. In some typical photoinduced excitonic dynamics processes in organic solar cells, the inter-state energy gaps and couplings sometimes may be larger[4, 151, 161-163]. Thus, two additional types of asymmetric models with $\Delta E = V_{12} = 0.1$ eV and $\Delta E = 2V_{12} = 0.2$ eV were used to check the performance of the MM-SQC method, see Figure 3 ($\Delta E = V_{12} = 0.1$ eV) and Figure 4 ($\Delta E = 2V_{12} = 0.2$ eV).

Due to the larger inter-state energy gap and the stronger electronic coupling, much faster system dynamics was observed compared to those of the first two types of the site-exciton models. Thus, we only provided the population evolution within the first 100 fs. For all models, the MM-SQC results are close to the ML-MCTDH ones, even when the high-frequency bath modes are treated by the classical manner. The good performance of the MM-SQC method here is attributed the fact that the fast electronic motion indicates that the bath motion becomes comparably slower in these models. The continuous rising of bath frequencies may lead to the inaccurate results of the MM-SQC dynamics, also due to the improper treatment of the fast bath modes. However, the bath with the very high characteristic frequency should not exist in some realistic situations, such as photoinduced exciton dynamics in organic photovoltaic systems. Only the stretching motions involving the H atoms, such as the OH, CH stretching motions, display such high vibrational frequency > 3000 cm$^{-1}$, while these motions normally display very small electron-phonon coupling in these processes[4], and should not play an essential role in photoinduced exciton dynamics in organic photovoltaic systems. The good performance of the SQC-MM dynamics in the description of the ultrafast exciton dynamics in organic semi-conducting polymer was recently discussed by Liang et al. [139]

Similar to the above stations, the triangle window method also performs a little better or gives equally good results in the current models with strong electronic couplings, with respect to the rectangle window method.

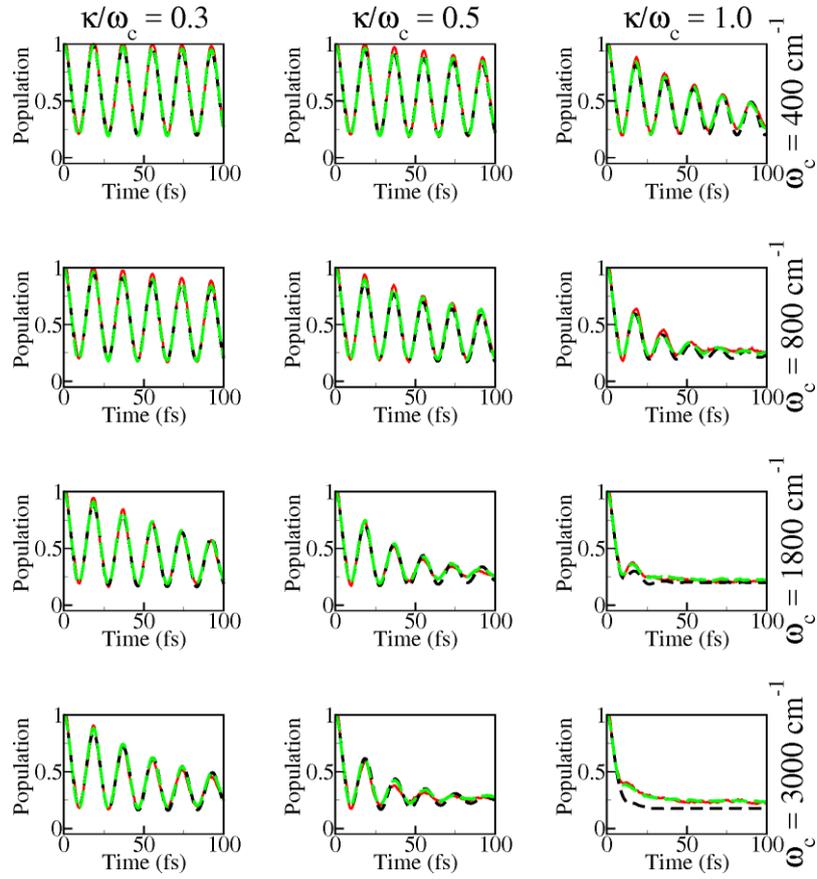

Figure 3. Time-dependent electronic population (or occupation) of the donor state in various two-state models with $\Delta E = V_{12} = 0.1$ eV. The characteristic bath frequencies $\omega_c$ are 400, 800, 1800 and 3000 cm$^{-1}$ and the electron-phonon coupling strength of the baths, $\kappa/\omega_c$, are 0.3, 0.5 and 1.0. Black dashed lines denote the ML-MCTDH results. Red solid lines and green dashed lines denote the MM-SQC results using rectangle window and triangle window, respectively.

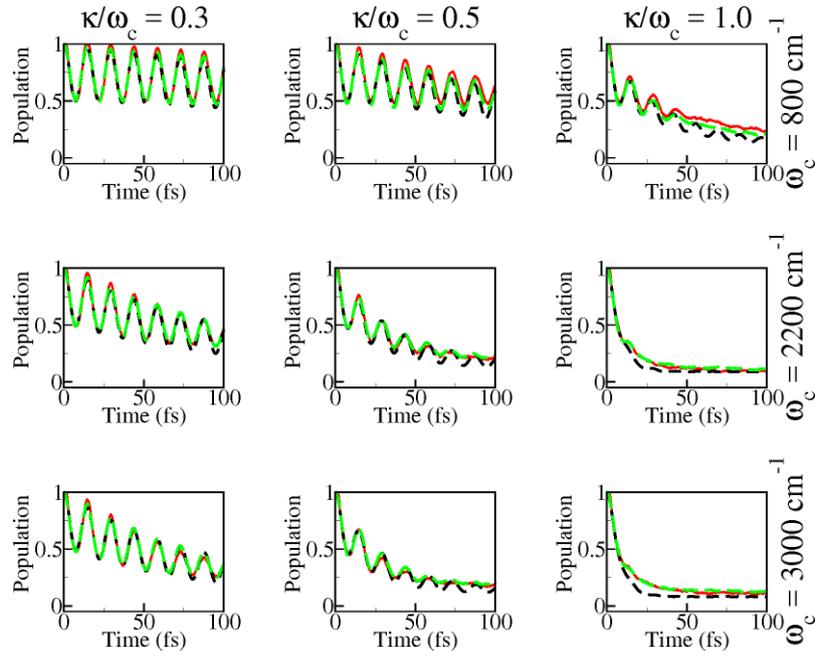

Figure 4. Time-dependent electronic population (or occupation) of the donor state in various two-state models with $\Delta E$ = 0.2 eV and $V_{12}$ = 0.1 eV. The characteristic bath frequencies $\omega_c$ are 800, 2200 and 3000 cm$^{-1}$ and the electron-phonon coupling strength of the baths, $\kappa/\omega_c$, are 0.3, 0.5 and 1.0. Black dashed lines denote the ML-MCTDH results. Red solid lines and green dashed lines denote the MM-SQC results using rectangle window and triangle window, respectively.

### 3.4 Initial Sampling of the Nuclear Configurations

The MM mapping procedure involves the transformation between the quantum occupation number of the harmonic oscillator and the physical variables (coordinate and momentum), which can also be written in the action-angle format. Thus, it is rather logical to employ the action-angle initial sampling for the mapping electronic degrees of freedom in the SQC-MM dynamics. However, for the nuclear degrees of freedom, there is no strict rule to provide such the quantum-classical correspondence. As a result, different ways may be employed for the initial sampling of the nuclear degrees of freedom, which mimic the corresponding quantum distribution. We noticed that many works on the MM-SQC dynamics employed the Wigner sampling for the nuclear degrees of freedom [55, 56, 126, 127, 131, 133, 135, 137]. Along this line, some of our calculations also employed the Wigner sampling for the bath modes. Certainly, it is also possible to perform the action-angle sampling for the nuclear modes. We thus try to consider different ways of the initial samplings for the nuclear degrees of freedom, including the Wigner sampling, the action angle samplings with different windows and the ZPE correction.

For the above symmetric site-exciton models, we observed that the Wigner sampling and the action-angle sampling ($\gamma_{nuc}$ = 0.5) methods give very similar results in the MM-SQC dynamics, see Figure 5. Please notice that the current ZPE correction value $\gamma_{nuc}$ corresponds to the half value of the parameter used in the previous work by Stock and coworkers [117, 118], because we use $\gamma_{nuc}$ not $\gamma_{nuc}/2$ to define the ZPE correction. Only when the ZPE correction employs small value of $\gamma_{nuc}$ =0.2, the action-angle sampling gives slightly different results in the strong electronic-phonon coupling cases, particular for the baths with high characteristic frequencies. Still, because all bath modes are treated by the classical evolution in the SQC-MM dynamics, the correct nonadiabatic dynamics cannot be recovered for the models with fast bath. Overall, different initial sampling approaches employed here seem not improve the performance of the MM-SQC dynamics for the symmetrical site-exciton models.

For the asymmetric site-exciton models, we have seen that obvious different results exist between the MM-SQC and ML-MCTDH dynamics when the fast bath is considered. Thus, we mainly discuss this situation here by employing a typical example with $\omega_c$ = 1000 cm$^{-1}$ (Figure 6). In these cases, the Wigner sampling and the action-angle sampling with $\gamma_{nuc}$ = 0.5 for the nuclear degrees of freedom give very similar results. However, in the current asymmetric site-exciton models, the inclusion of the ZPE correction for the nuclear degrees of freedom in the initial sampling improves the performance of the MM-SQC dynamics. The simulation results are dependent on the $\gamma_{nuc}$ value. The lower $\gamma_{nuc}$ values seem to give the lower donor-state population, particularly for the long-time dynamics. We also noticed that the lower ZPE correction $\gamma_{nuc}$ value should be employed for the strong electronic-phonon coupling cases.

According to the above observations, we realized that the influence of the initial sampling for the nuclear degrees of freedom is rather complicated. The MM-SQC dynamics seems not significantly depend on whether the Wigner sampling or the action angle sampling without the ZPE correction is used in the initial sampling of nuclear degrees of freedom. When the ZPE correction is employed in the action-angle sampling of nuclear part, different results are found in symmetric and asymmetric models. The ZPE correction in the initial sampling of the nuclear degrees of freedom seems not give the very deep impact on the overall dynamics in the symmetric site-exciton models, while the ZPE correction with different $\gamma_{nuc}$ value substantially changes the MM-SQC dynamics in the asymmetric site-exciton models.

Overall, the dependence of the nonadiabatic dynamics on the ZPE correction for the initial sampling of the nuclear degrees of freedom seems to be highly dependent on the system models. The current study mainly focuses on the performance of the MM-SQC dynamics, thus the deep understanding of the ZPE corrections is beyond the scope of this work. In fact, the research on the ZPE correction in the classical simulation is an extremely challenging topic[117, 118, 165-168].

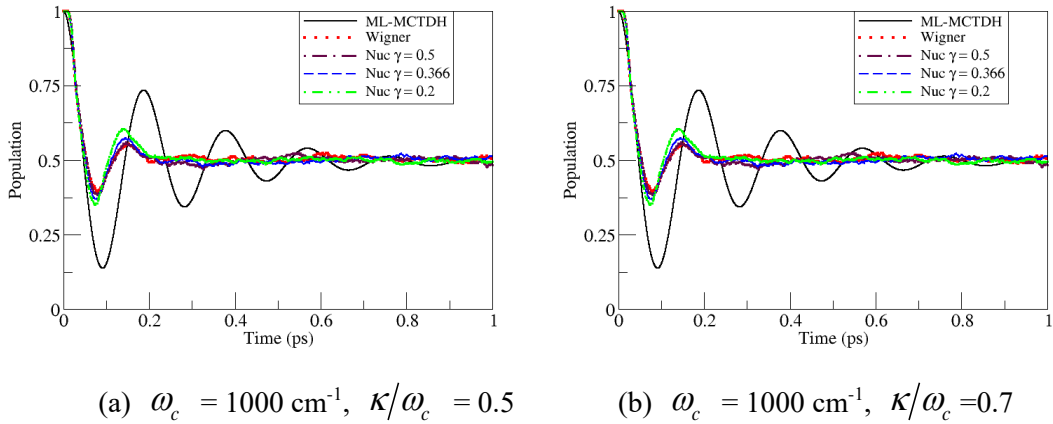

(a) $\omega_c = 1000$ cm$^{-1}$, $\kappa/\omega_c = 0.5$  (b) $\omega_c = 1000$ cm$^{-1}$, $\kappa/\omega_c = 0.7$

Figure 5. Time-dependent electronic population (or occupation) of the donor state in a two-state model with $\Delta E = 0$ and $V_{12} = 0.0124$ eV. Here the label "Nuc" means the ZPE correction is employed to in the initial action-angle sampling of the nuclear degrees of freedom.

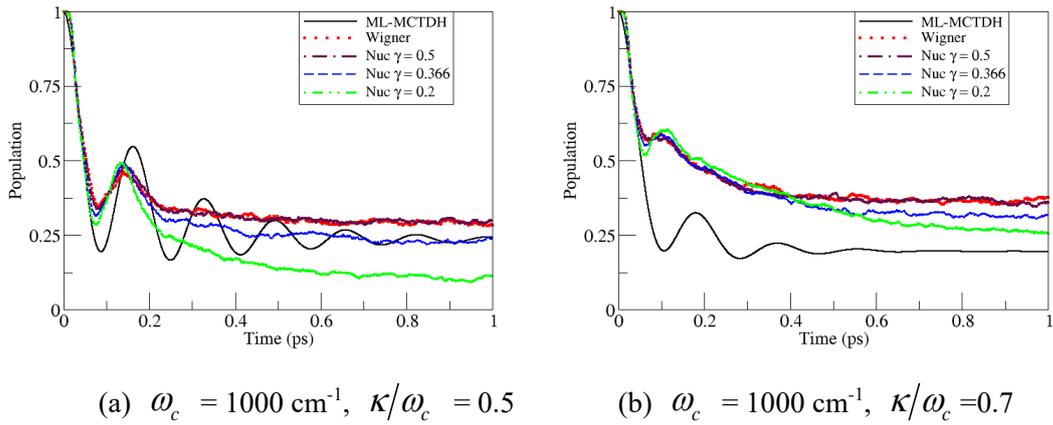

(a) $\omega_c = 1000$ cm$^{-1}$, $\kappa/\omega_c = 0.5$    (b) $\omega_c = 1000$ cm$^{-1}$, $\kappa/\omega_c = 0.7$

Figure 6. Time-dependent electronic population (or occupation) of the donor state in a two-state model with $\Delta E = V_{12} = 0.0124$ eV. Here the label "Nuc" means the ZPE correction is employed to in the initial action-angle sampling of the nuclear degrees of freedom.

## 3.5 Treatment of fast bath with adiabatic renormalization

Actually, the poor results by the MM-SQC method in the fast bath models are only relevant to the improper treatment of the high-frequency bath modes in the classical manner. In this situation, somehow it is meaningless to discuss the performance of the MM-SQC method. In principle, the high-frequency modes should be frozen in the quantum dynamics. Along this line, we took the so-called "adiabatic renormalization" approach [152] to remedy the problem when the high-frequency modes are involved. This idea was also widely employed in previous works[154-158]. In fact, some ideas in this treatment may be loosely relevant to the approach proposed by Cao and Geva [131].

After the employment of the adiabatic renormalization, both ML-MCTDH and MM-SQC dynamics simulations were performed using the new reduced model with the rescaled interstate coupling and only slow bath modes. The triangle window technique was employed in the MM-SQC dynamics. The results are shown in Figures 7 and 8 for symmetric and asymmetric models. The ML-MCTDH simulations using the reduced models produced by the adiabatic renormalization gave the same results as those of the original ones. It means these high-frequency modes are basically frozen at their ground vibrational levels in the dynamics of the current models. The MM-SQC dynamics simulations based on the reduced model also provided results similar to the ML-MCTDH ones. After this comparison, we can draw a conclusion that the MM-SQC method itself can always provide reasonable results in all test cases, while the proper treatments of the bath modes are very critical.

At the end, the current work chose the nuclear initial condition as the lowest vibrational level of the electronic ground state and then put it into the donor state. This choice was employed in several previous studies [4, 139, 169]. It is also preferable to use such initial condition in the ML-MCTDH treatment. We already demonstrated that the MM-SQC method performs quite well in the current initial conditions for all models, when the bath is treated properly. For example, the MM-SQC dynamics still works well for the fast bath modes, after the employment of the adiabatic renormalization. In this sense, these findings already provide a reasonable conclusion of the current work. Certainly, the other initial conditions, such as the non-zero temperature situations, are also important, and this should be a very interesting research topic in the future.

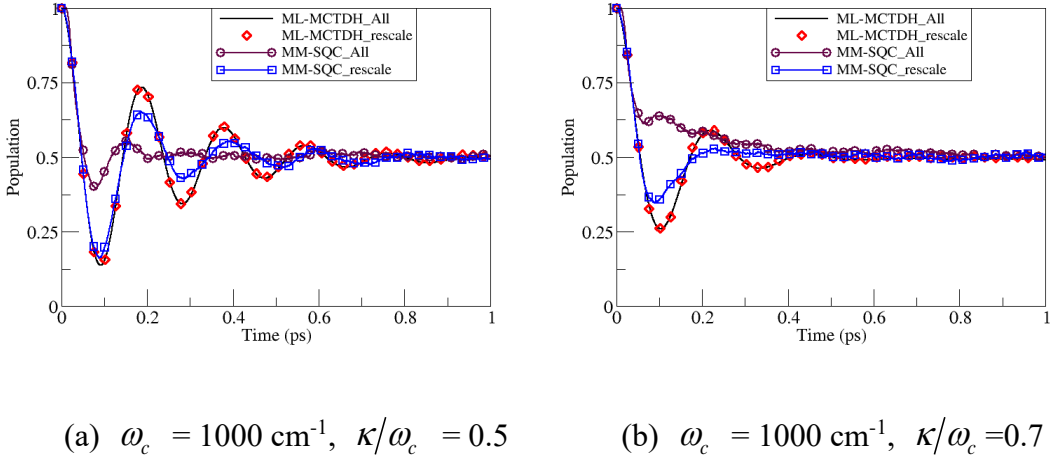

(a) $\omega_c = 1000$ cm$^{-1}$, $\kappa/\omega_c = 0.5$      (b) $\omega_c = 1000$ cm$^{-1}$, $\kappa/\omega_c = 0.7$

Figure 7. Time-dependent electronic population (or occupation) of the donor state in a two-state model with $\Delta E = 0$ and $V_{12} = 0.0124$ eV. Here the label "all" means that all bath modes are treated explicitly by the ML-MCTDH and MM-SQC methods. The label "rescale" means that the dynamics is performed based on the reduced model (with rescaled interstate coupling) produced by the adiabatic renormalization.

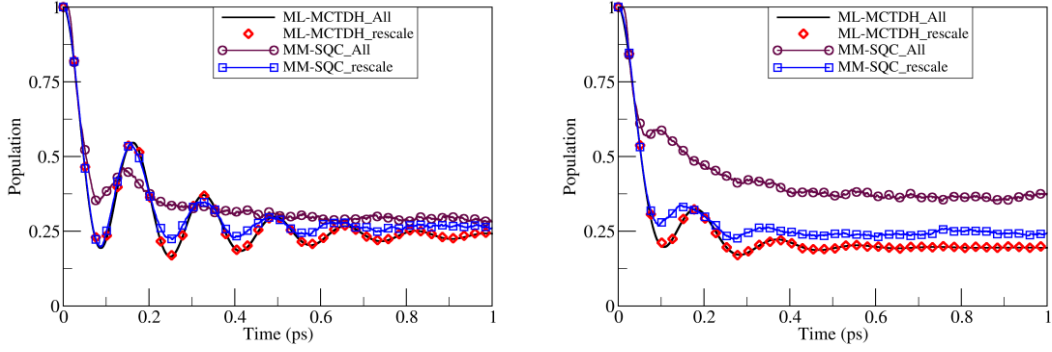

(a) $\omega_c = 1000$ cm$^{-1}$, $\kappa/\omega_c = 0.5$   (b) $\omega_c = 1000$ cm$^{-1}$, $\kappa/\omega_c = 0.7$

Figure 8. Time-dependent electronic population (or occupation) of the donor state in a two-state model with $\Delta E = V_{12} = 0.0124$ eV. Here the label "all" means that all bath modes are treated explicitly by the ML-MCTDH and MM-SQC methods. The label "rescale" means that the dynamics is performed based on the reduced model (with rescaled interstate coupling) produced by the adiabatic renormalization.

## 4. CONCLUSION

The performance of the MM-SQC method was extensively examined by employing four types of the site-exciton models representing the photoinduced charge/energy transfer processes. These models consider different Hamiltonian parameters, such as energy gap, electronic coupling, the bath characteristic frequency and electron-phonon coupling. The ML-MCTDH method was used to benchmark the results obtained from the MM-SQC dynamics. For the first two types of the site-exciton models with weak electronic coupling, 0.0124 eV, the MM-SQC method performs very well in the low-characteristic-frequency and weak electron-phonon-coupling regions, and it even provides the identical results as those of the ML-MCTDH method, even when all bath modes are treated classically. With the increasing of the characteristic frequency and electron-phonon coupling, the MM-SQC may not reproduce the ML-MCTDH results if the fast bath modes are still treated classically. In the third and fourth groups of models strong electronic coupling 0.1 eV is considered and the electronic motion becomes faster, the "fast bath" in the first two groups of the site-exciton models becomes the comparably slow in the last two groups of the site-exciton models with strong electronic couplings. As the result, the performance of MM-SQC method improves significantly in many models, even if the classical treatment is employed for all bath modes. However, this treatment may still bring problems when the characteristic frequency of the bath is extremely high and the electron-phonon couplings are very strong.

The dependence of the MM-SQC dynamics on the initial sampling of the nuclear degrees of freedom is also addressed. It is important to noticed that the Winger sampling and action angle sampling without the ZPE correction give the very similar results, no matter whether the symmetric or asymmetric site-exciton models are considered. In the action angle sampling, we still miss the comprehensive understanding of the possible effects of the ZPE correction for the nuclear degrees of freedom on the overall nonadiabatic dynamics. Different $\gamma_{nuc}$ value gives rather different population dynamics in the asymmetric site-exciton models, while such dependence seems to be minor in the symmetric site-exciton models. Thus, more further efforts should be devoted to understand the ZPE problem of the nuclear degrees of freedom in the MM-SQC dynamics.

As expected, it is not proper to treat the high-frequency bath modes purely classically, when the MM-SQC dynamics is employed. To provide a proper treatment, the adiabatic renormalization approach was employed to build reduced models, in which the fast modes (with frequency much large than the diabatic couplings) are frozen in the ground vibrational level of each diabatic state, only the low-frequency modes are treated explicitly and the diabatic coupling is rescaled. By applying the adiabatic renormalization, the MM-SQC dynamics can give the very good results compared to the exact ML-MCTDH dynamics of the full model. Thus, The MM-SQC method itself provides reasonable results in all current models, while the proper treatment of the bath modes is necessary.

Overall, the MM-SQC method performs well in our testing site-exciton models corresponding to photoinduced exciton processes in photoharvesting systems and organic solar cells. When the slow bath or the weak system-bath interactions are involved, this method gives very accurate results. When involving fast bath, this method also performs well with proper treatments by the employment of the adiabatic renormalization approach. Thus, the MM-SQC method is quite promising to treat the nonadiabatic dynamics of the complex systems within the the similar parameter regions discussed in this work. At the same time, it should be also highly interesting to combine it with the on-the-fly dynamics[170]. This represents the great challenging in the future.

## AUTHOR INFORMATION

**Corresponding Author**

\* E-mail: lanzg@qibebt.ac.cn; zhenggang.lan@gmail.com

Fax: +86-532-80662778; Tel: +86-532-80662630.

**Notes**

The authors declare no competing financial interest.

# ACKNOWLEDGEMENTS

This work is supported by NSFC projects (Nos. 21673266, 21503248 and 11747170). J. Z. thanks the Natural Science Foundation of Shandong Province (ZR2018BB043) and the Postdoctoral Scientific Research Foundation of Qingdao (2017012). The authors thank the Supercomputing Center; Computer Network Information Center, CAS; National Supercomputing Center in Shenzhen; National Supercomputing Center in Guangzhou and the Supercomputing Center of CAS-QIBEBT for providing computational resources.

# APPENDIX：

**I. Rectangle and Triangle Window Techniques**

In order to provide a clear comparison between two window techniques, we calculated the absolute population difference between the ML-MCTDH result and the MM-SQC result with the rectangle (or triangle) window technique, as shown in Figures 9 and 10. The performance of the MM-SQC dynamics with the triangle window technique is better than that with the rectangle window technique, especially in the early stage of the dynamics.

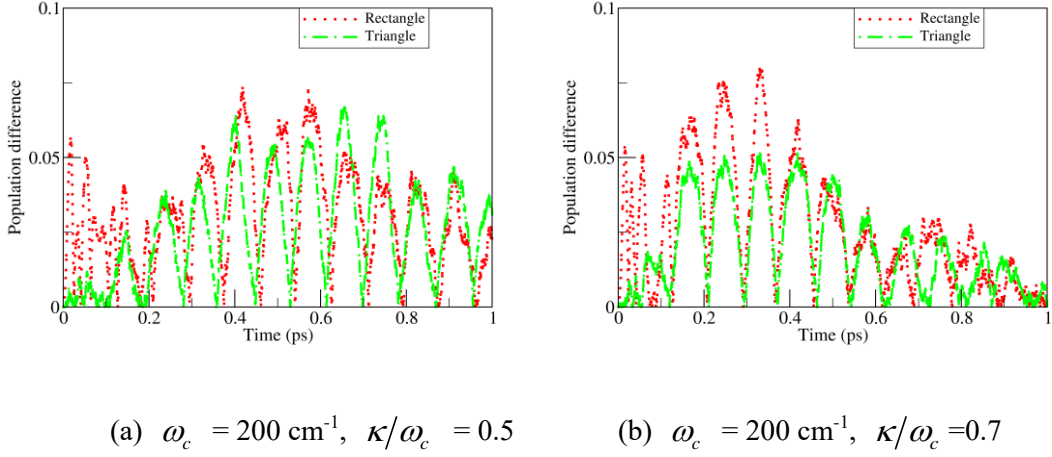

(a) $\omega_c = 200$ cm$^{-1}$, $\kappa/\omega_c = 0.5$      (b) $\omega_c = 200$ cm$^{-1}$, $\kappa/\omega_c = 0.7$

Figure 9. Time-dependent donor-state population (or occupation) difference between the ML-MCTDH method and the MM-SQC method using different windowing technique in various two-state models with $\Delta E = 0$ and $V_{12} = 0.0124$ eV.

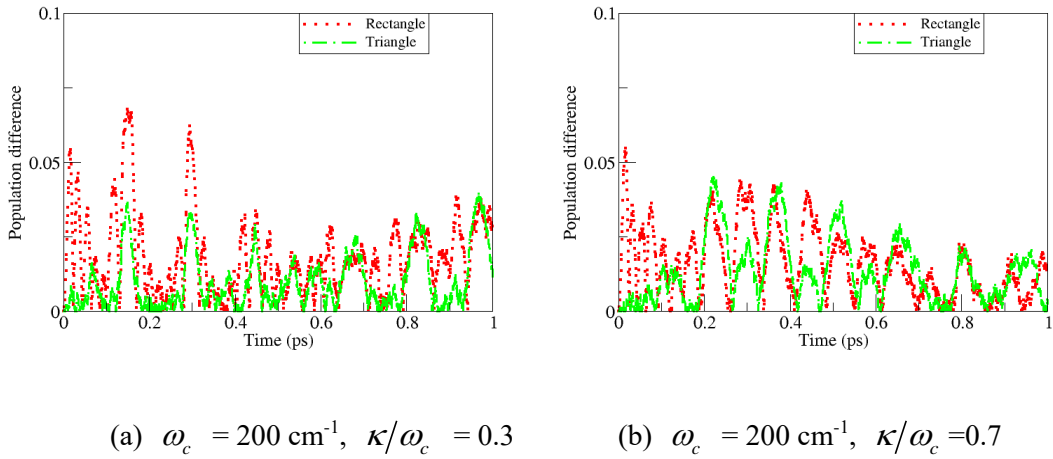

(a) $\omega_c = 200$ cm$^{-1}$, $\kappa/\omega_c = 0.3$      (b) $\omega_c = 200$ cm$^{-1}$, $\kappa/\omega_c = 0.7$

Figure 10. Time-dependent donor-state population (or occupation) difference between the ML-MCTDH method and the MM-SQC method using different windowing technique in various two-state models with $\Delta E = V_{12} = 0.0124$ eV.